\documentclass[aps,prl,amssymb,showpacs,twocolumn]{revtex4}
\usepackage{graphicx}

\begin{document}

\title{X-ray Radiation from the Annihilation of Dark Matter at the Galactic Center}

\author{Lars Bergstr\"om}
\email{lbe@physto.se}
\author{Malcolm Fairbairn}
\email{malc@physto.se}
\affiliation{Cosmology, Particle astrophysics and String theory, Department of Physics, Stockholm University, AlbaNova University Centre, SE-106 91, Stockholm, Sweden}
\author{Lidia Pieri}
\email{lidia.pieri@oapd.inaf.it}
\affiliation{INAF - Astronomical Observatory of Padova, Vicolo dell'Osservatorio 5, I - 35122 Padova, Italy}
\affiliation{INFN - Sezione di Padova}
\pacs{95.35.+d,11.10.Kk,98.70.Qy,98.35.Jk}

\newcommand{\lP}{\ell_{\mathrm P}}

\newcommand{\md}{{\mathrm{d}}}
\newcommand{\Kern}{\mathop{\mathrm{ker}}}
\newcommand{\tr}{\mathop{\mathrm{tr}}}
\newcommand{\sgn}{\mathop{\mathrm{sgn}}}

\newcommand*{\R}{{\mathbb R}}
\newcommand*{\N}{{\mathbb N}}
\newcommand*{\Z}{{\mathbb Z}}
\newcommand*{\Q}{{\mathbb Q}}
\newcommand*{\C}{{\mathbb C}}

\begin{abstract}

The existing and upcoming multiwavelength data from the 
Galactic Center suggest a comparative study in order to propose or rule out
possible models which would explain the observations. In this paper we
consider the X-ray synchrotron and the gamma-ray emission 
due to Kaluza Klein Dark Matter and define a set of parameters for the shape
of the Dark Matter halo which is consistent with the observations.  We show 
that for this class of models the existing Chandra X-ray data is more 
restrictive than the constraints on very high energy gamma-rays coming from HESS.

\end{abstract}

\maketitle

\section{Introduction}

Combined data from the cosmic microwave background radiation, the distant type 1a supernovae and from large scale structure studies suggest that approximately 25\%
 of the density content of the universe is non-baryonic dark matter (dark matter) \cite{spe:03}.  Weakly interacting massive particles (WIMPs) such as stable neutralinos in supersymmetric extensions of the standard model (SUSY) \cite{jun:96} or Kaluza Klein (KK) particles in theories where there is a TeV$^{-1}$ size universal extra dimension into which all standard model fields propagate \cite{app:01,servanttait} are exciting candidates since they can freeze out leaving a similar relic abundance to what is observed.   

Neutralino or KK particles annihilation into standard model particles may be detected 
astrophysically \cite{Bergstrom:2000pn} by observations of the regions of the Universe where the dark matter is expected to be densest. N-body simulations suggest that dark matter haloes may have  
cuspy density profiles with large density peaks in the core. It is thought that the center of the Milky Way
might contain such a dark matter overdensity, but it is also clear from observations that the gravitational field in the very central region is dominated by 
the $2.87 \pm 0.15 \times 10^6\ M_\odot$ supermassive black hole \cite{BH} which resides there.
The products of any dark matter annihilations will therefore be 
injected into the plasma falling into the central black hole.  In this work we calculate the signal expected from the synchrotron radiation due to the annihilation of KK particles into electrons and positrons near the Galactic Center. 

Dark matter annihilation products depend upon the type of candidate involved.
%Lidia
For instance, direct annihilation of SUSY dark matter into light fermions is highly suppressed and only low energy secondary electrons and positrons can arise as annihilation products. 
In KK scenarios there are universal extra dimensions with TeV$^{-1}$ size into which the standard model fields propagate and there is an orbifold condition which renders the lightest KK mode stable \cite{servanttait}.  In this model, the lightest KK mode is well approximated by the first KK mode of the B component of the electroweak field and therefore couples to standard model fermions and not gauge bosons (although photons may be produced via Bremsstrahlung processes \cite{gaugeboys}).
The annihilation into light fermions is no longer helicity suppressed and one might expect hard electrons from direct annihilation.

Since the electrons produced in the annihilation of SUSY dark matter typically have rather low energies compared to those produced in KK dark matter annihilation, their synchrotron radiation has a correspondingly smaller frequency.  Synchrotron from the electrons arising from the annihilation of SUSY WIMPs (studied in detail in reference \cite{olintoblasi}) would emerge at a frequency where there is either a lot of emission from the infalling gas or in a frequency band where one would expect a large amount of extinction.  In contrast, synchrotron from the hard electrons produced by the annihilation of KK particles will emerge at much higher frequencies and peaks close to the region of sensitivity of the Chandra X-ray telescope, which conveniently has extremely good angular resolution.  The background emission from Sagittarius A* is also lower in this region of frequency space, increasing the possibility of detection of a signal from the dark matter.  We will therefore concentrate on these hard electrons from KK WIMP annihilation, rather than the electrons from SUSY WIMP annihilation (we refer the reader to reference \cite{olintoblasi} for a detailed analysis of the latter).

\section{Galactic center}
\begin{figure}
\hspace{2.5cm}
\begin{center}
\includegraphics[height=8cm,width=6cm,angle=270]{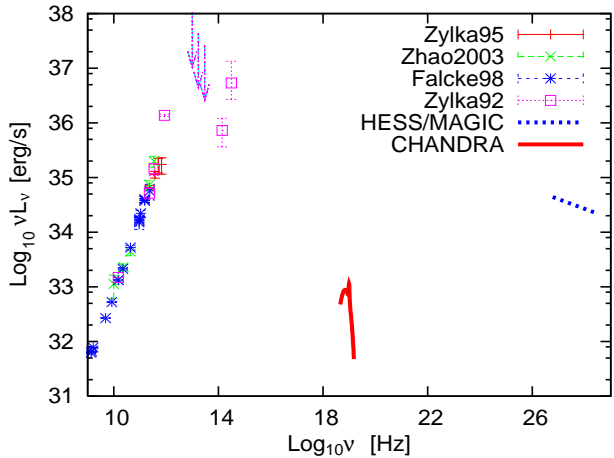} 
\caption{\it 
Multiwavelength luminosity of Sgr A* in the quiescent state. Observations in the radio and infrared come from referenes~\cite{zylka95,falcke98,zhao2003,zylka92}.  The Chandra data is reconstructed from reference~\cite{ironline}. The HESS/MAGIC spectrum is plotted as one line since the two experiments are in agrement with each other~\cite{hess,MagicTeV}.}
\label{flux}
\end{center}
\end{figure}
Multi-wavelength observations of the galactic centre, which we review in Fig.~\ref{flux} 
\footnote{The Chandra data were reconstructed assuming an effective area, defined as the product of the detection efficiency times the geometrical area of the detector of 400 $cm^2$}, have led to various models to explain the observed emission from the central black hole.  The consensus at the time of writing is that the sub-Eddington accretion flow onto the black hole is fueled by stellar mass loss from the cluster of large mass stars which exists in that region \cite{melia,melialiu,yuan}. 
The radio, mm and infrared radiation is thought to come 
from the inner regions of this flow close to the black hole, whereas 
the X-ray emission observed by Chandra is thought to originate further 
from the black hole, close to the Bondi radius at the interface between 
the spherical inflow region and the stellar winds where the gravity of 
the black hole starts to dominate the dynamics of the gas  \cite{baganoff2003,ironline}.  The Bondi radius is thought to be at around 0.04 pc from the central black hole, rather close to the 1 arcsecond resolution of the Chandra telescope at this distance.
HESS observations \cite{hess} which have recently been confirmed by the MAGIC experiment \cite{MagicTeV} show that there is significant TeV gamma-ray emission from the central 30 pc around the black hole.  This emission might be due to the annihilation of dark matter \cite{horns,gaugeboys,fps:04,profumo} or might have a more mundane origin, being created by Fermi acceleration in shock fronts in the stellar winds \cite{fermi}, or as a product of the interaction of ultra high energy protons with ambient photons and magnetic field, or as initiated by proton - proton interactions in the accretion disk, or generated by curvature and inverse Compton emission of accelerated electrons close to the Black Hole \cite{aha-ner:05}.  Recently it has been pointed out that this emission may hinder searches for the annihilation of dark matter because it provides too great a background \cite{hooper}.

\begin{figure}
\begin{center}
\includegraphics[height=8cm,width=9cm]{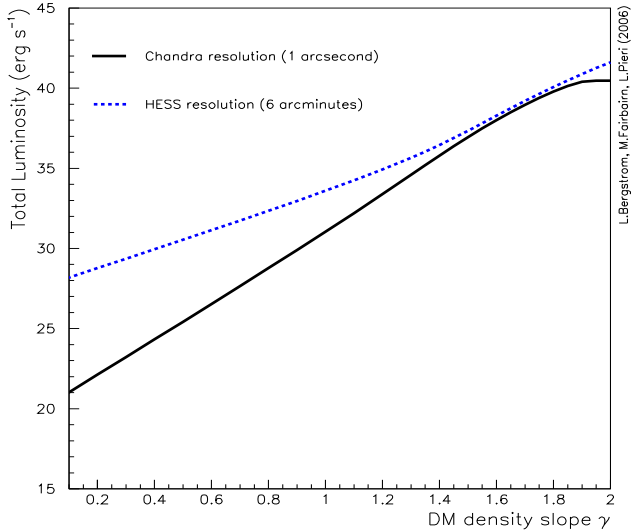} 
\caption{\it Expected total luminosity from the center of the galaxy for different dark matter profiles of the form $\rho\propto r^{-\gamma}$ normalised to $0.3$ GeV cm$^{-3}$ at the sun, for $\langle\sigma_{tot} v\rangle = 3 \times
10^{-26} \ cm^3 \ s^{-1}$ and $m_{dm} = 1 \ TeV$. Since the emission is diffuse, the two curves correspond to 6 arcminutes (HESS) and arcsecond (Chandra) angular resolution.  We assume a maximum density in the core due to self annihilation (see text).  The luminosity stops growing for the Chandra case when the entire region within the Chandra angular resolution is at the maximum density of dark matter.}
\label{lumt}
\end{center}
\end{figure}

In order to calculate the expected luminosity coming from the annihilation of WIMPS, one first needs to know what the density profile is within the region in question.  Many N-body simulations predict that the density at the center of dark matter halos will asymptote to a power law $\rho\propto r^{-\gamma}$ \cite{mooreold,NFWold,1.18} so the simplest approach is to assume a simple power law and to normalise it so that the local density at the sun is $0.3$ GeV cm$^{-3}$.  Assuming the emission along the line of sight is dominated by the galactic central region, the luminosity expected from that region is given by
%Lidia
\begin{equation}
L=f_{em}\langle\sigma_{tot} v\rangle m_{dm}4\pi\int_{r_{min}}^{r_{max}}\left(\frac{\rho(r)}{m_{dm}}\right)^2r^2 dr
\label{lumintegral}
\end{equation}
where $f_{em} \sim 0.5$ is the fraction of all the final states, like electrons, muons, taus and quark jets that will give rise to electromagnetic energy and  $\langle\sigma_{tot} v\rangle$ is the total thermally averaged KK particle annihilation cross
section.  The inner radius $r_{min}$ is the cut-off radius below which there is a maximum density core due to the high self-annihilation rate.  If we assume that the dark matter halo has existed for a time $\tau_h$ then then the radius $r_{min}$ is defined by $\rho(r_{min})=m_{dm}\langle\sigma_{tot} v\rangle/\tau_h$.  The outer radius $r_{max}$ corresponds to the angular resolution of the instrument in question at the distance corresponding to the centre of the galaxy. This may be an underestimate since the contribution from the dark matter at $r>r_{max}$ along the line-of-sight is not considered. However, in our models, which have rather steep central densities, the assumption that the dark matter emission emanates entirely from a sphere with radius corresponding to the angular resolution of the instrument is a good approximation.  Eq. \ref{lumintegral} can therefore be considered a correct estimate.
In figure \ref{lumt} we give the luminosity expected to lie within the beam of an arcminute resolution device such as HESS or GLAST and an arcsecond resolution telescope such as Chandra.

Typical values of the asymptotic power law for the density profile in the inner regions found in N-body simulations are $\gamma\sim 1-1.5$. The quiescent X-ray emission observed by Chandra is around $10^{33}$ erg s$^{-1}$, which is interesting as it is rather close to the emission that one would expect from the annihilation of WIMPs from the same region for these values of $\gamma$.

Generation of energetic electrons in a magnetized plasma will lead to synchrotron radiation and in order to predict the synchrotron spectrum, we need a model for the magnetic field around the center of the galaxy.

\section{Magnetic Field}

Radio observations of the galactic centre show some evidence of variability thought to be associated with a very small central accretion disk at a scale of $2.7\times 10^{-4}$ arcseconds, i.e. 40 $R_{BH}$ \cite{Doeleman:2001nr}, where $R_{BH} \sim 7 \times 10^{11} \ cm$ 
is the Schwarzschild radius for the black hole.
At much larger radii it is therefore valid to assume the behaviour of the magnetic field is governed by the general theory of quasi-spherical infall. Observations over many years also show that, despite flaring, accretion onto Sagittarius A* appears to be quasi-static over time (see e.g. \cite{Herrnstein:2004ue}).

The theory of quasi-static infalling plasma predicts a magnetic field profile of the form $B(r) \sim r^{-2}$. This power law holds until some radius $r_{equi}$ at which equipartition between magnetic and kinetic energy is achieved.
\begin{equation}
\frac{\rho(r) v(r)^2}{2} = \frac{B^2(r)}{8 \pi}
\end{equation}
For radii smaller than $r_{equi}$, accretion is possible only if the magnetic
field is destroyed and the magnetic energy dissipated. There are strong constraints on the approximate strength of the magnetic field in the central regions in order to agree with the observed sub-mm radiation \cite{yuan,melialiu}.

The simplest way to normalise the strength of the magnetic field in the GC is to assume that the inflowing gas is due to the stellar outflow from large stars in the center of the galaxy \cite{melia}.  This gives rise to a mass inflow at the accretion radius, which sets the electron density and equipartition magnetic fields from that radius downwards.

In our model we have $r_{equi} \sim 0.04 \rm pc$ and $\dot{M}=10^{22} g s^{-1}$ and if we assume standard spherical Bondi-Hoyle accretion below that radius, we then have an equipartition magnetic field of strength 
\begin{equation}
B_{eq}(r)=3.9\times 10^{-2} \left(\frac{0.01{\rm pc}}{r}\right)^{\frac{5}{4}}\rm Gauss
\label{equipartition}
\end{equation}
in agreement with the authors of \cite{olintoblasi}.

\begin{figure}[t!]
\begin{center}
\includegraphics[height=8cm,width=9cm]{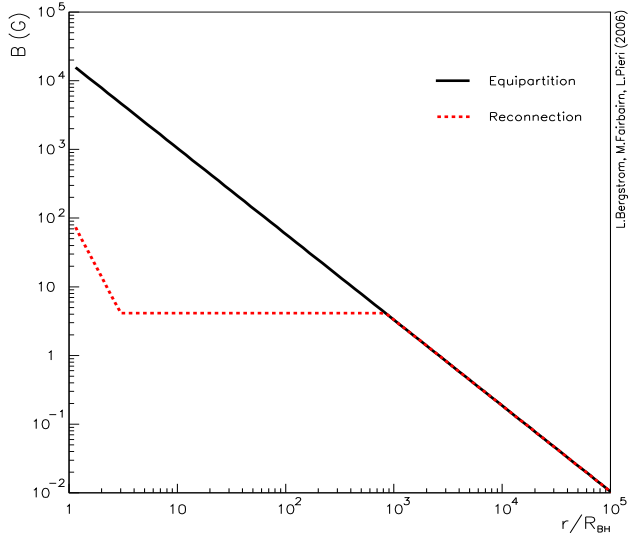} 
\caption{\it Assumed magnetic field as a function of radius from the Black Hole. The solid line corresponds to the equipartition magnetic field of Eq.~\ref{equipartition} whereas the red dashed line is the flux-reconnection magnetic field of Eq.~\ref{bmelia}.}
\label{bfield}
\end{center}
\end{figure}

It has been pointed out however that the equipartition picture may not be correct at very small distances from the GC, where magnetic field line reconnection in the turbulent plasma may reduce the magnetic field. We will therefore also calculate the spectrum corresponding to the following modified magnetic field:
\begin{eqnarray}
B_m(r) = B_{eq}(r) &&  r > 10^3 R_{BH} \nonumber \\
B_m(r) = B_{eq}(10^3 R_{BH}) && 3 R_{BH} < r < 10^3 R_{BH} \nonumber \\
B_m(r) = B_{eq}(3 R_{BH})   \left (\frac{r}{3 R_{BH}}\right)^{-3} && R_{BH} < r < 3 R_{BH}
\label{bmelia}
\end{eqnarray}
Here the various length scales have been taken from the re-connected field in reference \cite{cokermelia}, although the overall strength of that field is slightly lower at large radii than the naive equipartition value which we adopt.  The two magnetic fields are plotted in figure \ref{bfield}. To a first approximation, it turns out that the synchrotron spectrum arising from the annihilation of dark matter in our region of interest ($\sim 10^{19} - 10^{20} \ Hz$) is not particularly sensitive as to whether one chooses the B-field described by equation  (\ref{equipartition}) or (\ref{bmelia}), at least in the window of sensitivity of Chandra.

There is rather a lot of uncertainty in the values of the magnetic fields which should be adopted in this very central region of the galaxy but we will show that the spectra in the region of interest are not extremely sensitive to the magnetic field.

\section{Energy loss mechanisms}

Since the synchrotron lifetime of TeV electrons in this environment is much shorter than the time scales of any of the other energy loss mechanisms, we can neglect them in the solution to the diffusion loss equation.  For example, at these high energies radiative processes dominate and energy losses for an electron with energy $E$ via a radiative mechanism $X$ can be expressed in the form
\begin{equation}
\dot{E}=\frac{4}{3}c\sigma_T U_X\gamma^2
\end{equation}
where $\gamma$ is the Lorentz factor and $U_X$ represents the energy density contained in the magnetic field for synchrotron losses, the background radiation field for inverse compton scattering (ICS) or the synchrotron radiation itself for synchrotron self comptonisation (SSC).  If we consider ICS then the CMB will contribute 0.25 eV cm$^{-3}$ whereas normal stellar radiation will contribute a few eV cm$^{-3}$ \cite{olintoblasi}.  With the fields that we have assumed the magnetic field $B>10^{-2}$ G in the region of interest which corresponds to $U_{sync} \propto B^2 \geq 10^6$ eV cm$^{-3}$,much larger than $U_{ics}$ in the central accretion region.
While subdominant, the ICS radiation will emerge at much higher energies than X-ray, for example the CMB photons will be scattered to tens of MeV, the region soon to be probed by the GLAST mission. 

One can show that the SSC is subdominant through numerical integration over the synchrotron flux to obtain $U_{ssc}$.  A simpler, rougher way of seeing this is by checking that the (over-)estimate of $U_{ssc}$ obeys the inequality
\begin{equation}
U_{ssc}\simeq\frac{Lr}{4\pi r^3 c}\ll \frac{B^2}{8\pi} = U_{sync}
\end{equation}
where $L$ is the luminosity of the system, $r$ is the radius and $c$ is the speed of light. This is simply the amount of synchrotron radiation that one would expect to flow through a unit volume at a given radius due to the total luminosity at smaller radii.  Since the dark matter profiles we consider are rather steep, most of the emission will come from smaller radii rather than from larger radii which is why this is a good approximation to the full integral over all radii to obtain the background synchrotron photons.

 In the results that we present below it turns out that only for the reconnection magnetic field and the most spiked profiles (profile C below) could this inequality be in danger, and then only in a small region of the emission region close to the black hole. Since we will see that such profiles are already ruled out by gamma-ray observations, we conclude that we do not need to worry about synchrotron-self absorption at the level of accuracy of this paper.

Other timescales are also larger than the energy loss time scale.  For example, the gravitational infall timescale compared with the synchrotron timescale for a TeV electron in the equipartition field (\ref{equipartition}) is given by
\begin{eqnarray}
\tau_{grav}=\sqrt{\frac{r^3}{GM_{BH}}}=2.7\times 10^8\left(\frac{r}{\rm 0.01pc}\right)^{\frac{3}{2}}\rm sec\nonumber\\
\tau_{sync}=\frac{3}{4\sigma_T c}\frac{8\pi}{B^2}\frac{E}{\gamma^2}=2.6\times 10^5\left(\frac{r}{\rm 0.01pc}\right)^{\frac{5}{2}}\rm sec\nonumber\\
\end{eqnarray}
so that $\tau_{grav}\gg\tau_{sync}$, demonstrating our point.  The characteristic timescale upon which the synchrotron timescale itself varies is very close to the gravitational timescale.  

We assume that the electrons lose energy before they change position significantly.  In order for this to be true, the diffusion length scale should be much smaller than the radial distance of the electron from the central black hole.  We obtain the diffusion length scale in the same way as the authors of reference \cite{bertonesync} by taking the geometric average of the magnetic diffusion length scale $d_{B}$ (taken to be one third of the gyromagnetic radius) with the distance corresponding to the synchrotron lifetime $c\tau_{sync}$.

\begin{equation}
\frac{\sqrt{c\tau_{sync}d_{B}}}{r}=\frac{m_e}{r}\sqrt{\frac{2\pi}{e\sigma_TB^3}}=2.78\times 10^{-4}\left(\frac{r}{\rm 0.01 pc}\right)^{\frac{7}{8}}
\end{equation}
so that the diffusion of the electrons can also be neglected.

We will therefore assume that all terms other than synchrotron energy loss can be set to zero, which would certainly not be true for electrons arising from SUSY WIMP annihilations.

When considering direct dark matter annihilation into electrons 
we will be interested in a delta function of electrons with energy $m_{dm}$.
The solution of the diffusion-loss equation then has the following form
%Lidia
\begin{equation}
\frac{dn}{dE}(E,r)=\frac{1}{2}\left(\frac{\rho(r)}{m_{dm}}\right)^2
\langle\sigma_{tot} v\rangle N_{ee}b_{ee}\frac{1}{\dot{E}}{\rm cm^{-3}GeV^{-1}}
\end{equation}
which is valid over a range of energies, $E<m_{dm}$. $N_{ee} = 2$ 
is the overall number of electrons and positrons produced in
each annihilation and $b_{ee} = 0.19$ is the branching ratio of
annihilation in the electron-positron line. We
used $\langle\sigma_{tot} v\rangle = 3 \times 10^{-26}$cm$^3$s$^{-1}$.  
$N(E,r)$ is zero for $E>m_{dm}$.

\section{Dark Matter Density Profiles}

The dark matter density profile, $\rho(r)$, at the GC is a subject of rich debate.
N-body simulations predict a density profile with an asymptotic power law behavior in the central region, the density rising as $\rho\propto r^{-\gamma}$ with different predictions for $\gamma$ ranging between 1 and 1.5 \cite{NFWold,mooreold,1.18}.  Other predictions suggest that the value of $\gamma$ changes steadily as one approaches the center of the halo \cite{improved}.

At the same time, it is well known that baryons in gravitationally bound systems lose energy and fall to the center, creating an enhanced potential well into which the dark matter is then drawn \cite{blumenthal}.  This phenomenon of adiabatic contraction is not completely understood, especially in the very central regions of the galaxy, although attempts to take into account the non-circularity of the orbits seem to help \cite{gnedin}.

It has also been suggested that the profile of dark matter in the immediate vicinity of the GC is enhanced during the period of formation of the black hole \cite{gondoloandsilk}.  This would lead to a rather dense ``spike'' of dark matter at the center which would in turn give rise to large annihilation rates.  However, more recently it has been argued that the dark matter in this spike would be heated by the gravitational dynamical friction of the stellar population, leading somewhat to its dispersion \cite{merrittbertone}.  We will use the results of the numerical work of reference \cite{merrittbertone} to obtain hopefully realistic models of the dark matter profile near the center of the galaxy.

\begin{figure}[t!]
\begin{center}
\includegraphics[height=8cm,width=9cm]{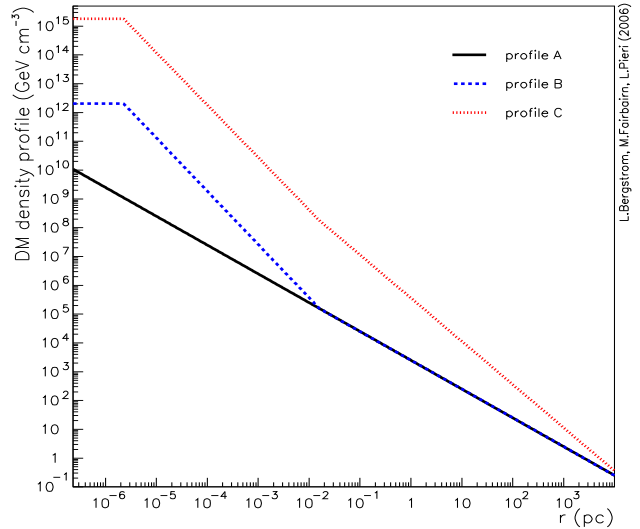} 
\caption{\it Density profiles used in this work and presented in table \ref{proftab}. }
\label{profpic}
\end{center}
\end{figure}

In light of the above discussion, we consider three density profiles for the dark matter distribution, all of which can be parametrised rather simply by the following expression 
\begin{eqnarray}
&\rho(r)=\rho(100\rm pc)\left(\frac{100\rm pc}{r}\right)^{\gamma_1}\qquad & r > r_{out}\nonumber\\
&\rho(r)=\rho(r_{out})\left(\frac{r_{out}}{r}\right)^{\gamma_2}\qquad & r_{out}> r> r_{in}\nonumber\\
&\rho(r)=\rho(r_{in})\qquad & r_{in} > r
\label{profile}
\end{eqnarray}
The three models we will consider are: A) the standard NFW $\gamma=1$ profile with no adiabatic contraction or central spike.  B) The same $\gamma=1$ profile, but now with a central spike which has diffused away over time, considerably reducing its density.  C) A profile which has undergone adiabatic contraction on galactic scales due to the presence of baryons, and also has a central spike, also allowed to diffuse away over time.  The profiles are summarised in table \ref{proftab} and plotted in figure \ref{profpic}.

For profile A) $r_{out}$=$r_{in}$=0 and $\rho(100\rm pc)=25\rm GeV/cm^3$ where the value of the density is obtained by normalising to the canonical density 0.3$\rm GeV/cm^3$ at the solar radius.  The parameters for profiles B) and C) are obtained by making approximate fits to the results published in figure 1 of reference \cite{merrittbertone}.  For both profiles $r_{out}=7\times 10^4 \ R_{BH}$ and $r_{in}=10 \ R_{BH}$.

\begin{table}
\caption{Parameters of the density profiles (see equation (\ref{profile})).  They are approximations of the profiles presented in \cite{merrittbertone}.\label{proftab}}
\begin{ruledtabular}
\begin{tabular}{cccccc}
Profile&$\rho(100\rm pc)$&$\gamma_{1}$&$\gamma_2$&$r_{out}$&$r_{in}$\\
\hline
A)&$25\ \rm GeV/cm^3$& 1 & - & 0&0\\
B)&$25\ \rm GeV/cm^3$& 1 & 1.85 &$7\times 10^4 \ R_{BH}$&$10 \ R_{BH}$\\
C)&$360\ \rm GeV/cm^3$& 1.5 & 1.82 &$7\times 10^4 \ R_{BH}$&$10 \ R_{BH}$\\
\end{tabular}
\end{ruledtabular}
\end{table}

\section{Results}

The X-ray synchrotron flux emitted in the central region of the Galaxy is
attenuated mainly via photoelectric absorption, which is the dominant process
for X-ray absorption up to energies of at least 100 keV.  This effect depends weakly upon atomic arrangement and can be computed considering free atoms.  The magnitude of the effect therefore depends upon the column density of electrons along the
line of sight $N_H$, which we set equal to $1.5 \times 10^{23} \ cm^{-2}$~\cite{porquet:03} and on the photoelectric cross section $\sigma_{p.e.}$~\cite{matt}.  

Because we ignore all energy loss effects other than synchrotron the equations for emissivity and luminosity simplify considerably.  We define $r_{res}$ to be the radius corresponding to the angular resolution of Chandra and $f(\nu',\nu)$ the fraction of synchrotron radiation emitted at a frequency $\nu'$ from an electron circling with a syncrotron frequency $\nu$.  The observed luminosity in synchrotron radiation of frequency $\nu'$ is then well approximated by the expression
\begin{eqnarray}
\nu'L_{\nu'}&=&\frac{1}{2}
\langle\sigma_{tot} v\rangle N_{ee}b_{ee}m_e \int_0^{r_{res}} 4\pi r^2\left(\frac{\rho(r)}{m_{dm}}\right)^2 \nonumber \\
&\times&\exp(-N_H\sigma_{p.e.}(\nu'))\int_0^\infty  f(\nu',\nu)\nonumber\\
&\times&\theta\left(1-\frac{\nu_*}{B_*(r)}\frac{m_{e}}{m_{dm}}\right)\sqrt{\frac{\nu_*}{B_*(r)}} d\nu dr
\end{eqnarray}
where the dimensionless quantity $\nu_*$ is $\nu/$Hz, $B_*=B/2.8\times 10^{-6}$ G and $\theta$ is the heavyside step function.  The function $f(\nu',\nu)$ is defined to be
\begin{equation}
f(\nu',\nu)=x\int_x^\infty K_{5/3}(y)dy\hspace{0.5cm},\hspace{0.5cm} x=\frac{\nu'}{\nu}
\end{equation}
where $K_{5/3}$ is a modified Bessel function.

\begin{figure}[t!]
\begin{center}
\includegraphics[height=8cm,width=9cm]{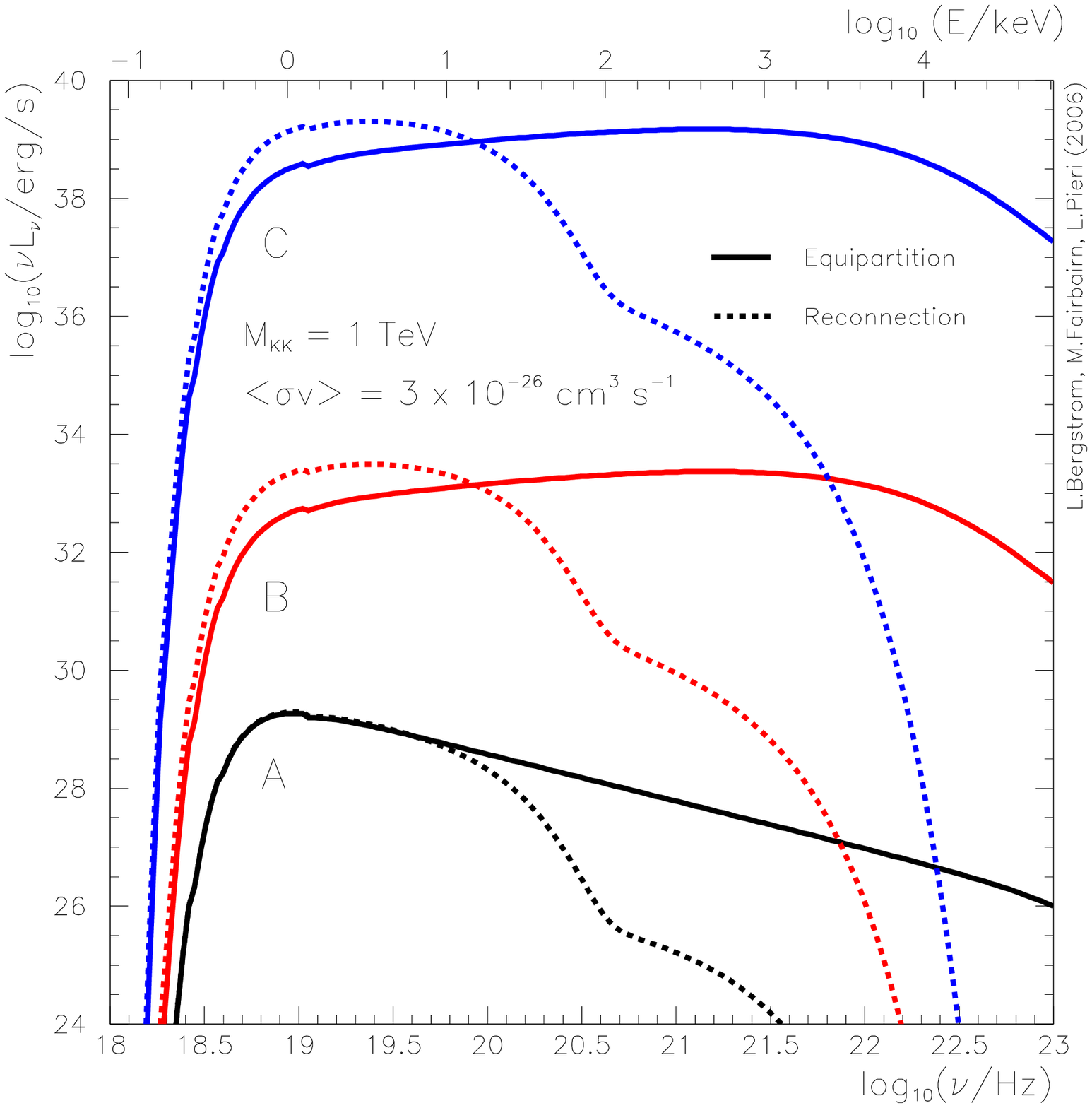} 
\caption{\it Synchrotron spectra from 1 TeV KK dark matter annihilation in the central 0.01 pc of the galaxy assuming the three density profiles (A,B,C) described in Eq.~\ref{profile}. The solid lines correspond to the equipartition magnetic field of Eq.~\ref{equipartition} and the dotted lines are the spectra with the flux-reconnection magnetic field of Eq.~\ref{bmelia}.}
\label{spectra}
\end{center}
\end{figure}

Figure \ref{spectra} shows the synchrotron spectra from a 1 TeV KK particle annihilation in the GC, for the different density profiles described in Eq.~\ref{profile}
and for the different magnetic field assumption of Eqs.~\ref{equipartition} 
and~\ref{bmelia}.

In order to find out if the X-ray emission predicted in our model is reasonable, we have to compare it with the HESS data to make sure that the haloes we consider do not give rise to too much emission in gamma-rays.  First we assume that the HESS resolution corresponds to a 30 pc radius sphere around the GC \cite{hess}, then we note that the authors of \cite{gaugeboys} fit the HESS data with an NFW $\gamma=1$ profile and a boost factor of 200 in the flux.  It is therefore necessary to ensure that the profiles that we use are not so dense as to saturate this bound, otherwise one would expect more emission in the form of TeV gamma-rays than observed by HESS.  The HESS bound corresponds to a total luminosity from within the 30 pc sphere of about $6.9\times 10^{37}\rm GeV s^{-1}$ whereas the three profiles A), B) and C) that we have considered correspond to $3.5\times 10^{30}, 5.5\times 10^{35}$ and $1.3\times 10^{41} \rm GeV s^{-1}$ respectively, so that profile C is ruled out.  

\begin{figure}
\begin{center}
\includegraphics[height=8cm,width=9cm]{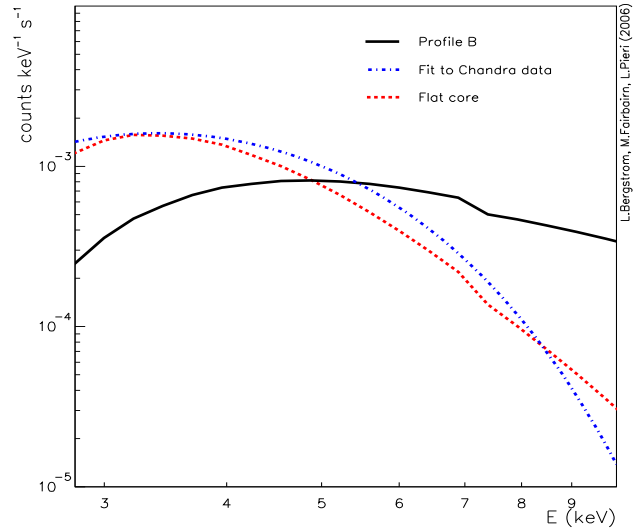} 
\caption{\it Comparison of profile B with Chandra data: the dot-dashed curve is an approximate fit to the data presented in \cite{baganoff2003} without the iron line.  The solid curve is the signal expected from synchrotron radiation from electrons produced in dark matter annihilations assuming density profile B.  The dashed curve corresponds to the synchrotron radiation from the flat core described in the text. We assume a Chandra effective aperture of 400 cm$^2$.}
\label{match}
\end{center}
\end{figure}

Profile B, which does not violate the bound from HESS, gives rise to approximately the same flux as the observed signal from Chandra in the region of interest as can be seen in figure \ref{match}.  In this way one can claim that X-ray observations are therefore more restrictive than TeV observations, since they rule out density profiles which are less steep than those ruled out by HESS.

It would be tempting to argue that the observed emission in X-ray could be explained via dark matter synchrotron rather than thermal-bremsstrahlung.  As we see in figure \ref{lumt}, the flux from dark matter synchrotron is certainly conceivably of the right order of magnitude, although the spectrum seems to have the wrong shape given the magnetic fields considered in this work.  There is also the observation of an iron line \cite{ironline}, a spectral feature which could not be explained very easily by synchrotron.  

It is interesting to see if it is possible to fit the continuum emission observed by Chandra and assumed to be thermal bremsstrahlung.  The observed spectrum drops more rapidly than our synchrotron, so we need to assume a magnetic field of the form ($\ref{bmelia}$) but with a strength one order of magnitude smaller than that plotted in figure \ref{bfield}, so that its maximum is at a lower energy than the absorption cut off at 2 keV.  If we then assume a core of dark matter with constant density of around $10^8$ M$_\odot$ pc$^{-3}$ then we can obtain a spectrum rather close to what is observed.

While this is an amusing result, it would be rather optimistic to claim that the continuum component of X-rays observed at the galactic center comes from the synchrotron radiation associated with dark matter electrons.  Nevertheless, we feel it is important to note that the energy injected into the plasma in the form of electrons is significant compared to the energy emitted in the X-ray region of the spectrum.  A more detailed study of the effect of these electrons as they thermalise and heat the local environment might be worthwhile.

\begin{figure}
\begin{center}
\includegraphics[height=8cm,width=9cm,angle=0]{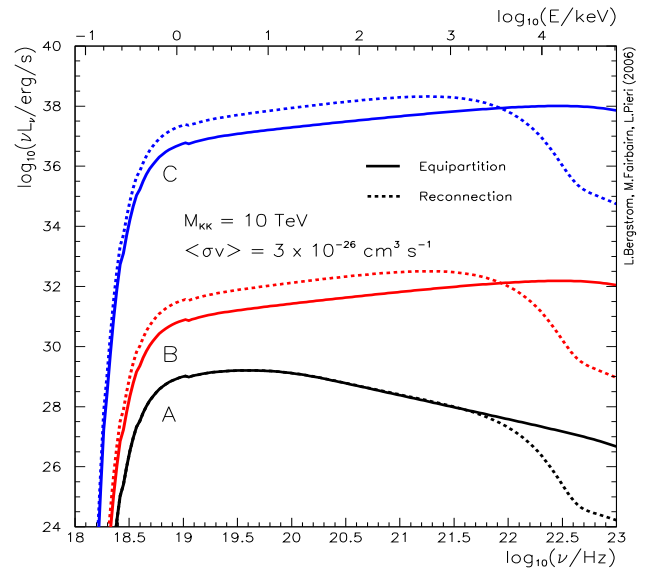} 
\caption{\it Synchrotron spectra from 10 TeV KK dark matter annihilation in the central 0.01 pc of the galaxy assuming the three density profiles (A,B,C) described in Eq.~\ref{profile}. The solid lines correspond to the equipartition magnetic field of Eq.~\ref{equipartition} and the dotted lines are the spectra with the flux-reconnection magnetic field of Eq.~\ref{bmelia}.}
\label{10teV}
\end{center}
\end{figure}

To be consistent with the studies \cite{gaugeboys,profumo,horns}, we can also consider the spectra emitted for different masses of KK particle.  This is perhaps not so interesting for the case of universal extra dimensions, because it is only when the KK particle has a mass close to 1 TeV that one obtains a good relic abundance.  However, for the sake of completeness we have calculated the spectra for a 10 TeV particle which annihilates into electrons the same branching ratio as KK particles in figure \ref{10teV}.

We have therefore presented the expected spectra from dark matter for three density profiles which seem to well motivated from astrophysical considerations at the time of writing.  A more general approach to the density profiles, which may be more appropriate given the large amount of uncertainties involved in their derivation, is the following.  We assume that the density of dark matter at the solar radius is 0.3 GeV cm$^{-3}$ and then we choose a single power law that is valid down to very small radii.  This is clearly unrealistic at the very center of the galaxy due to the dynamics discussed in the previous section, but it does serve as a useful parametrization.  

We find that the steepest profile which is compatible with the X-ray data is $r^{-1.35}$.  The gamma rays produced by such a profile within the angular resolution of the HESS telescope array are much less than what is observed.  Consequently we find that the X-ray observations from Chandra can be much more restrictive than the data from gamma ray telescopes with much larger angular uncertainties.

\section{Conclusions}

In this paper we have calculated the expected X-ray synchrotron spectra flux from high energy electrons produced by the annihilation of KK dark matter particles at the galactic center.  Many of our conclusions will be approximately valid for other TeV dark matter candidates which decay into hard fermions without helicity suppression.

We presented spectra for two different magnetic fields, one corresponding to equipartition with the plasma falling into the central black hole, the second taking into account the possibility of flux reconnection which may occur due to turbulence in the infalling gas.  We also looked at three different density profiled, showing how they affected the expected spectra.

The luminosity expected from the galactic center region due to the annihilation of WIMPS is rather close to what is actually observed (within a few orders of magnitude either way, depending upon the assumed density profile.)  This is remarkable because of the physics which governs the flux from dark matter annihilations is completely different to that governing the accretion onto the central black hole.  The electrons injected into the plasma due to the annihilation of dark matter may therefore have a considerable effect upon the astrophysics of the central region around the black hole.

We found that the X-ray emission from the GC is not inconsistent with the annihilation of KK particles of mass 1 TeV, providing the shape of the inner density profile is less steep than $r^{-1.35}$.  Since the total luminosity corresponding to this profile within the angular resolution of the HESS telescope is less than the luminosity which has been observed by HESS, we are able to claim that X-ray observations from Chandra are more constrictive than existing gamma-ray data.  Because of this, it should be impossible to detect KK dark matter using gamma-rays, since if it were possible, the X-ray synchrotron signal of that dark matter should already have produced much more flux in X-rays than what is observed in the Chandra data.  

{\bf Acknowledgments} LB, MF and LP are grateful for receiving funding from the Swedish Research Council (Vetenskapsr\aa det).

\end{document}